Relativistic five-quark equations and a narrow $N^*(1688)$ resonance.

Gerasyuta S.M.[1], Kochkin V.I.

1. Department of Physics, LTA, 194021 St. Petersburg, Russia,

gerasyuta@sg6488.spb.edu

Abstract.

The relativistic five-quark equations are found in the framework of the dispersion relation technique. The five-quark amplitude for the $N^*(1688)$ resonance including $u, d$ quarks is calculated. The pole of this amplitude determines the mass of non-strange pentaquark M=1686 MeV and the width about 7 MeV. The results are in good agreement with the evidence for a narrow $N^*(1688)$ resonance.

I. Introduction

The experimental study of the quasi-free $\gamma n \to \eta n$ reaction at GRALL [1], CB/TAPS [2] and LNS-Tohoku [3] facilities provided the evidence for a relatively narrow resonant structure at M=1.67-1.68 GeV. The structure has been observed as a bump in the quasi-free cross section and in the $\eta n$ invariant mass spectrum. Reportedly, its decay width was estimated to be about 40 MeV. However the Fermi motion being taken into account the width may be found to be around 10 MeV [4]. Such bump is not seen in $\eta$ photoproduction on the proton. The cross section on the proton exhibits only a minor peculiarity in this mass region [5]. Therefore the bump in $\eta$ photoproduction on the neutron may be a manifestation of a neutron resonance with unusual properties: the possibly narrow width and the much stronger photocoupling to the neutron than to the proton. Its identification is now a challenge for both theory [6-10] and experiment [11, 12].

The evidence for a narrow resonant structure in the beam asymmetry data for $\eta$ photoproduction on the free proton was reported. This structure may be a manifestation of a narrow resonance with the mass M=$1.688 \pm 0.002 \pm 0.005$ GeV and the width $\Gamma \leq 15$ MeV.

In Refs. [13-17], a method developed which is convenient for analyzing relativistic three-hadron systems. The physics of three-hadron systems can be described by means of a pair interaction between the particles. There are three isobar channels, each of which consists of a two-particle isobar and the third particle. The presence of the isobar representation together with the condition of unitarity in the pair energies and of analyticity leads to a system of integral equations in a single variable. Their solution makes it possible to describe the interaction of the produced particles in a three-hadron system.

In our papers [18, 19] the relativistic generalization of the three-body Faddeev equations was obtained in the form of dispersion relations in the pair energy of two interacting particles. The mass spectrum of S-wave baryons including u, d and s quarks was calculated by the method based on isolating the leading singularities in the amplitude.

We searched for the approximate solution of integral three-quark equations by taking into account two-particle and triangle singularities, all the weaker ones being neglected. If we considered such an approximation, which corresponds to taking into account two-body triangle singularities and defined all smooth functions of the subenergy variables (as compared with the singular part of the amplitude) in the middle point of the physical region of Daltz-plot, then the problems was reduced to the one of solving a system of simple algebraic equations.

In this paper the relativistic generalization of five-quark equations (like Faddeev – Yakubovsky approach) are used in the form of the dispersion relation [20-22]. The five-quark amplitude for the $N^*(1688)$ pentaquark contains the $u$-, $d$- quarks. The pole of this amplitude determines the mass of $N^*(1688)$ M=1686 MeV and the width $\Gamma = 7$ MeV. The masses of the constituent $u$-, $d$- quarks m=410 MeV coincide with the quark masses of the ordinary baryons in our quark model [19].

The paper is organized as follows. After this introduction, we discuss the five-quark amplitudes which contain $\bar{u}$ - antiquark and two pair of $u$-, $d$- quarks. In the section 3, we report our numerical results.

## II. Five-quark amplitude for $N^*(1688)$ baryon.

We derived the relativistic five-quark equations in the framework of the dispersion relation technique [23]. The correct equations for the amplitude are obtained by taking into account all possible subamplitudes. It corresponds to the division of complete system into subsystems with a smaller number of particles. Then one should a five-particle amplitude as a sum of ten subamplitudes:

$$A = A_{12} + A_{13} + A_{14} + A_{15} + A_{23} + A_{24} + A_{25} + A_{34} + A_{35} + A_{45}. \tag{1}$$

This defines the division of the diagrams into group according to certain pair interaction of particles. The total amplitude can be represented as a sum of diagrams. We need to consider only one group of diagrams and the amplitude corresponding to them, for example $A_{12}$. We shall consider the derivation of the relativistic generalization of the Faddeev-Yakubovsky approach for $N^*(1688)$ nonstrange pentaquark. We shall construct the five-quark amplitude using $u$-, $d$- quarks, in which the pair interactions with the quantum numbers of a $J^P = 0^+$ diquark is included. The pole of this amplitude determines the masses of $udud\bar{u}$ and $udud\bar{d}$ pentaquarks.

In order to represent the subamplitudes $A_1(s, s_{1234}, s_{12}, s_{34})$, $A_2(s, s_{1234}, s_{25}, s_{34})$, $A_3(s, s_{1234}, s_{13}, s_{134})$, and $A_4(s, s_{1234}, s_{23}, s_{234})$ in the form of a dispersion relation it is necessary to define the amplitudes of quark-quark and quark-antiquark interaction $b_n(s_{ik})$. The pair quarks amplitudes $q\bar{q} \to q\bar{q}$ and $qq \to qq$ are calculated in the framework of the dispersion N/D method with the input four-fermion interaction [24-26] with quantum numbers of the gluon [27].

$$g_v(\bar{q}\lambda I_f \gamma_\mu q)^2 \qquad (2)$$

Here $I_f$ is the unity matrix in the flavor space ($u$, $d$), $\lambda$ are the color Gell-Mann matrices. Dimensionless parameters $g$ and $\lambda$ are supposed to be constants which are independent of the quark interaction type:

$$g = \frac{m^2}{\pi} g_v, \quad \lambda = \frac{\Lambda}{m^2} \qquad (3)$$

The applicability of Eq. (2) is verified by the success of De Rujula-Georgi-Glashow quark model [28], where only the short-range part of Breit potential connected with the gluon exchange is responsible for the mass splitting in hadron multiplets.

We use the results of our relativistic quark model and write down the pair quarks amplitude in the form:

$$b_n(s_{ik}) = \frac{G_n^2(s_{ik})}{1 - B_n(s_{ik})}, \qquad (4)$$

$$B_n(s_{ik}) = \int_{4m^2}^{\Lambda} \frac{ds'_{ik}}{\pi} \frac{\rho_n(s'_{ik}) G_n^2(s'_{ik})}{s'_{ik} - s_{ik}}. \qquad (5)$$

Here $s_{ik}$ is the two-particle subenergy squared, $s_{ijk}$ corresponds to the energy squared of particles $i$, $j$, $k$, $s_{ijkl}$ is the four-particle subenergy squared and $s$ is the system total energy squared. $G_n(s_{ik})$ are the quark-quark and quark-antiquark vertex functions. The vertex functions are determined by the contribution of the crossing channels. These vertex functions satisfy the Fierz relations. All of these vertex functions are generated from $g_v$. $B_n(s_{ik})$ and $\rho_n(s_{ik})$ are the Chew-Mandelstam functions with the cutoff $\Lambda$ and the phase spaces respectively.

In the case in question the interacting quarks do not produce a bound state; therefore the integration in Eqs.(6) - (9) below is carried out from the threshold $4m^2$ to the cutoff $\Lambda_n$. The system of integral equations, corresponding to the meson state with $J^{PC} = 0^{++}$ and diquark with $J^P = 0^+$, can be described as:

$$A_1(s, s_{1234}, s_{12}, s_{34}) = \frac{\lambda_1 B_2(s_{12}) B_1(s_{34})}{[1 - B_2(s_{12})][1 - B_1(s_{34})]} + 3\hat{J}_2(2,1) A_4(s, s_{1234}, s'_{23}, s'_{234}) +$$
$$+ 2\hat{J}_2(2,1) A_3(s, s_{1234}, s'_{13}, s'_{134}) + 2\hat{J}_1(2) A_3(s, s_{1234}, s'_{15}, s_{125}) + 2\hat{J}_1(2) A_4(s, s_{1234}, s'_{25}, s_{125}) +, \quad (6)$$
$$+ 2\hat{J}_1(1) A_4(s, s_{1234}, s'_{35}, s_{345})$$

$$A_2(s, s_{1234}, s_{25}, s_{34}) = \frac{\lambda_2 B_1(s_{25}) B_1(s_{34})}{[1 - B_1(s_{25})][1 - B_1(s_{34})]} + 6\hat{J}_2(1,1) A_4(s, s_{1234}, s'_{23}, s'_{234}) +$$
$$+ 8\hat{J}_1(1) A_3(s, s_{1234}, s'_{12}, s_{125}) \quad (7)$$

$$A_3(s, s_{1234}, s_{13}, s_{134}) = \frac{\lambda_3 B_2(s_{13})}{1 - B_2(s_{13})} + 8\hat{J}_3(2) A_1(s, s_{1234}, s'_{12}, s'_{34}), \quad (8)$$

$$A_4(s, s_{1234}, s_{23}, s_{234}) = \frac{\lambda_4 B_1(s_{23})}{1 - B_1(s_{23})} + 2\hat{J}_3(1) A_2(s, s_{1234}, s'_{25}, s'_{34}) + 2\hat{J}_3(1) A_1(s, s_{1234}, s'_{12}, s'_{34}), \quad (9)$$

were $\lambda_i$ are the current constants. Here n=1 corresponds to the $qq$-pair with $J^P = 0^+$ in the $\bar{3}_c$ color state. n=2 defines the $q\bar{q}$ pairs, corresponding to meson with quantum numbers $J^{PC} = 0^{++}$.

We introduced the integral operators:

$$\hat{J}_1(l) = \frac{G_l(s_{12})}{[1 - B_l(s_{12})]} \int_{4m^2}^{\Lambda} \frac{ds'_{12}}{\pi} \frac{G_l(s'_{12}) \rho_l(s'_{12})}{s'_{12} - s_{12}} \int_{-1}^{+1} \frac{dz_1}{2}, \quad (10)$$

$$\hat{J}_2(l, p) = \frac{G_l(s_{12}) G_p(s_{34})}{[1 - B_l(s_{12})][1 - B_p(s_{34})]} \times$$
$$\times \int_{4m^2}^{\Lambda} \frac{ds'_{12}}{\pi} \frac{G_l(s'_{12}) \rho_l(s'_{12})}{s'_{12} - s_{12}} \int_{4m^2}^{\Lambda} \frac{ds'_{34}}{\pi} \frac{G_p(s'_{34}) \rho_p(s'_{34})}{s'_{34} - s_{34}} \int_{-1}^{+1} \frac{dz_3}{2} \int_{-1}^{+1} \frac{dz_4}{2}, \quad (11)$$

$$\hat{J}_3(l) = \frac{G_l(s_{12}, \tilde{\Lambda})}{1 - B_l(s_{12}, \tilde{\Lambda})} \times$$

$$\times \frac{1}{4\pi} \int\limits_{4m^2}^{\tilde{\Lambda}} \frac{ds'_{12}}{\pi} \frac{G_l(s'_{12}, \tilde{\Lambda}) \rho_l(s'_{12})}{s'_{12} - s_{12}} \int\limits_{-1}^{+1} \frac{dz_1}{2} \int\limits_{-1}^{+1} dz \int\limits_{z_2^-}^{z_2^+} dz_2 \frac{1}{\sqrt{1 - z^2 - z_1^2 - z_2^2 + 2zz_1z_2}}, \qquad (12)$$

were $l, p$ are equal 1 or 2.

In Eqs.(10) and (12) $z_1$ is the cosine of the angle between the relative momentum of the particles 1 and 2 in the intermediate state and the momentum of the particle 3 in the final state, is taken in the c.m. of particles 1 and 2. In Eq.(12) $z$ is the cosine of the angle between the momenta of the particles 3 and 4 in the final state, taken in the c.m. of particles 1 and 2. $z_2$ is the cosine of the angle between the relative momentum of particles 1 and 2 in the intermediate state and the momentum of the particle 4 in the final state, taken in the c.m. of particles 1 and 2. In Eq. (11): $z_3$ is the cosine of the angle between relative momentum of particles 1 and 2 in the intermediate state and the relative momentum of particles 3 and 4 in the intermediate state, taken in the c.m. of particles 1 and 2. $z_4$ is the cosine of the angle between the relative momentum of the particles 3 and 4 in the intermediate state and momentum of the particle 1 in the intermediate state, taken in the c.m. of particles 3, 4.

We can pass from the integration over the cosines of the angles to the integration over the subenergies.

The solution of the system of equations is considered as:

$$\alpha_i(s) = F_i(s, \lambda_i) / D(s), \qquad (13)$$

where zero of $D(s)$ determinant defines the mass of bound state of pentaquark. $F_i(s, \lambda_i)$ are the functions of $s$ and $\lambda_i$. The functions $F_i(s, \lambda_i)$ determine the contributions of subamplitudes to the exotic baryon amplitude.

## III. Calculation results.

The functions $F_i(s, \lambda_i)$ allow to obtain the overlap factors $f$ for the pentaquarks $N^*(1688)$. We calculate the overlap factor $f$ and the phase spaces for the reaction $N^* \to \eta N$. Given the width of pentaquark $\theta^+(1540)$ about 1 MeV, we would estimate naively that the $N^*(1688)$ width is $\Gamma \approx f^2 \times \rho$ [29, 30]. We calculate the width $\Gamma = 7$ MeV, and mass M=1686 Mev for the pentaquark $N^*(1688)$ with isospin I=1/2 and $J^P = \frac{1}{2}^+$ ($udud\bar{u}$).

The model has only two parameters. The cutoff parameter $\Lambda_{0^+}$ and the gluon coupling constant $g = 0.417$ are similar to Ref. 23. Williams and Gueye have developed a quark molecular model (QMM) of strangeness S=0 pentaquarks, using a molecule structure, and predicted the low-lying mass spectrum [31]. They considered the molecules $uds - u\bar{s}$ and received the masses of about 1800-2000 MeV. We take into account only u, d- quarks and received smaller mass for the $N^*(1688)$.

## Acknowledgments.

The authors would like to thank T. Barnes, S. Chekanov and V. Kuznetsov for useful discussions. This work was carried with the support of the Russian Ministry of Education (grant 2.1.1.68.26).